\documentstyle[11pt,psfig]{article}
\begin{document}
\title{Counting surface states\\
in the loop quantum gravity}
\author{Kirill V.Krasnov\thanks{E-mail address: krasnov@phys.psu.edu} \\
\it Center for Gravitational Physics and Geometry, \\ 
\it The Pennsylvania State University, University Park, PA 16802, USA \\
\it and \\
\it Erwin Schr\"{o}dinger Institute for Mathematical Physics,\\  
\it Boltzmanngasse 9, 1090 Vienna, Austria \\ 
\it and \\
\it Bogolyubov Institute for Theoretical Physics, Kiev 143, Ukraine.}
\maketitle
\begin{abstract}
We adopt the point of view that (Riemannian) classical  and (loop-based) quantum descriptions of geometry are macro- and micro-descriptions in the usual statistical mechanical sense. This gives rise to the notion of geometrical entropy, which is defined as the logarithm of the number of different quantum states which correspond to one and the same classical geometry configuration (macro-state). We apply this idea to gravitational degrees of freedom induced on an arbitrarily chosen in space 2-dimensional surface. Considering an `ensemble' of particularly simple quantum states, we show that the geometrical entropy $S(A)$ corresponding to a macro-state specified by a total area $A$ of the surface is proportional to the area $S(A)=\alpha A$, with $\alpha$ being approximately equal to $1/16\pi l_p^2$. The result holds both for case of open and closed surfaces. We discuss briefly physical motivations for  our choice of the ensemble of quantum states.
\end{abstract}
\bigskip

\section{Introduction.}

Among the results of the loop approach to non-perturbative quantum 
gravity there are several which tell us that the 
picture of geometry on scales small as compared with our usual 
scales (on Planckian scales) looks quite differently from the 
habitual picture of Riemannian geometry. The usual classical 
picture seems to arise only on coarse-graining. The fundamental excitations of 
the emerging quantum geometry are one-dimensional loop excitations and the 
whole quantum picture is of essentially discrete, combinatorial 
character. (See, for instance, recent works \cite{QGeometry,GeomEig}, 
which also contain extensive references to the previous papers on the 
loop approach).

Let us assume now that we can formulate a reasonable approximation criteria, and for any classical 
geometry configuration find the set of approximating it quantum 
states.
It is natural to expect that if the precision of approximation is chosen 
not too high, there will 
be a lot of quantum states corresponding to the same `geometry'. It is
quite tempting to consider a usual Riemannian geometry
description as a macroscopical one and to regard all quantum states 
approximating a Riemannian metric as micro-states corresponding to the same
macro-state. This point of view brings up a lot of interesting 
possibilities.    

Indeed, recall that with distinguishing between macro- and micro-states of a system the notion of entropy arises in 
statistical mechanics. Entropy
is a function which depends on a macroscopic state of the system. It can be
thought of as a function which for each macroscopic state gives the (logarithm
of the) number of different microscopic states corresponding to this 
macro-state. More precisely, the space of states of the system should be divided into compartments, where all micro-states
belonging to the same compartment are macroscopically indistinguishable. Then
the entropy of a macro-state is given\footnote{%
In fact, the usual thermodynamical entropy is proportional to this number, but it is convenient to work in the units in 
which this proportionality coefficient is chosen to be unity.}
by the logarithm of the `volume' of the compartment corresponding to this 
macro-state. 

Let us return to our quantum description of geometry.
It is natural to introduce a coarse-graining of
the space of quantum states in such a way 
that quantum states approximating different
`geometries' belong to different compartments. Having divided the space of states
this way, it is natural to introduce the function which for any 
geometry configuration
gives the logarithm of the `volume' of the corresponding compartment. This
gives rise to the notion of {\it geometrical entropy}.
Thus, geometrical entropy tells `how many' there are different quantum
states which correspond to a given geometry configuration. To be more precise, in the cases
when a compartment is itself a linear space the entropy
is given by the logarithm of the dimension of the corresponding compartment.

In this paper we try to implement the idea of geometrical entropy following a very simple choice of the `ensemble' of quantum states. Although some of our results are surprising, the aim of this paper is not to argue a physical significance of the results obtained. Rather, in order to understand if the general idea of geometrical entropy makes sense, we consider a particularly simple case, in which the analysis can be accomplished, and develop a technique that may prove to be useful for future developments. 

Our choice of the `ensemble' of quantum states is as follows. 
First, we restrict our consideration to the gravitational degrees of freedom of an arbitrarily chosen in space
2-dimensional surface. Namely, we consider Lorentzian 3+1 general 
relativity in the framework of (real) Ashtekar variables, and the 
quantization given by the loop quantum gravity.
We restrict our consideration to the 
gravitational system induced by the full theory
on some 2-surface $S$ embedded in the spatial manifold $\Sigma$. Thus, the 
degrees of freedom of our system are just the 
degrees of freedom of general relativity that live on a surface $S$. Second, we specify a macro-state of our system simply fixing the total area of the surface $S$. In this case, as we shall see later, it is easy to pick up all quantum states which approximate a given macro-state, and the analysis becomes almost straightforward. Thus, to illustrate the general idea of geometrical entropy we stick here to the case when a macro-state of our system is specified simply by the total area $A$ of the surface.

Let us note that such a statistical mechanical consideration of surface degrees of freedom is of a special 
interest because of its possible connection with the black hole 
thermodynamics. Indeed, there is a common believe that it is the 
degrees of freedom living on the horizon surface of a black hole
which account for the black hole entropy. 
To try to reveal the connection 
between quantum gravity and thermodynamics that is suggested by 
black hole physics is one the motivations for our investigation.

The other, and maybe even more important motivation is 
that the loop quantum gravity itself is a new approach. In such
a situation it is necessary to apply the formalism to simple problems, just
in order to see if it gives a reasonable picture. The set of problems
concerning statistical properties of the theory might serve as one 
of such tests.

The paper is organized as follows. In the next section we remind 
briefly how the surface quantum states 
look like and discuss the issue of correspondence between macro- and micro-descriptions. In Section \ref{sec:3} we calculate geometrical entropy of surface degrees of freedom considering the case of an open surface. Section \ref{sec:4} contains a generalization of our result to the case of closed surfaces. We conclude with the discussion.

\section{Surface quantum states.}
\label{sec:2}

Let us recall the description of general relativity in terms of 
(real) Ashtekar variables. In the Hamiltonian framework general 
relativity can be formulated as a theory of $SU(2)$-connection 
over the spatial manifold. The connection field plays the role of a 
configurational variable; the momentum variable is presented by 
the canonically conjugated field. The dynamics of the theory is 
determined by a set of constraint functionals.

The degrees of freedom induced on an arbitrarily chosen surface $S$ 
are described by pull-backs of the connection and momentum fields 
into $S$, which we shall denote by  $a_a^{AB}$ and ${\tilde e}^{a\,AB}$ 
respectively ($A,B$ stands for two component spinor indices).
The `surface' momentum field ${\tilde e}^{a\,AB}$ carries information about 2-metric on $S$.

Let us describe the quantum theory. The loop quantization of 3+1 general relativity is described in details in 
\cite{Asht1}. For our purposes it is sufficient to  recall that there exists
an orthogonal decomposition of the Hilbert space of gauge invariant states of quantum general relativity into subspaces, which are labeled by the so-called spin network states. Spin network states are labeled by closed graphs in
$\Sigma$ with spins (or, equivalently, with irreducible representations of the gauge group) assigned to each edge and intertwining operators assigned to each vertex of graph.

\begin{figure}
\centerline{\hbox{\psfig{figure=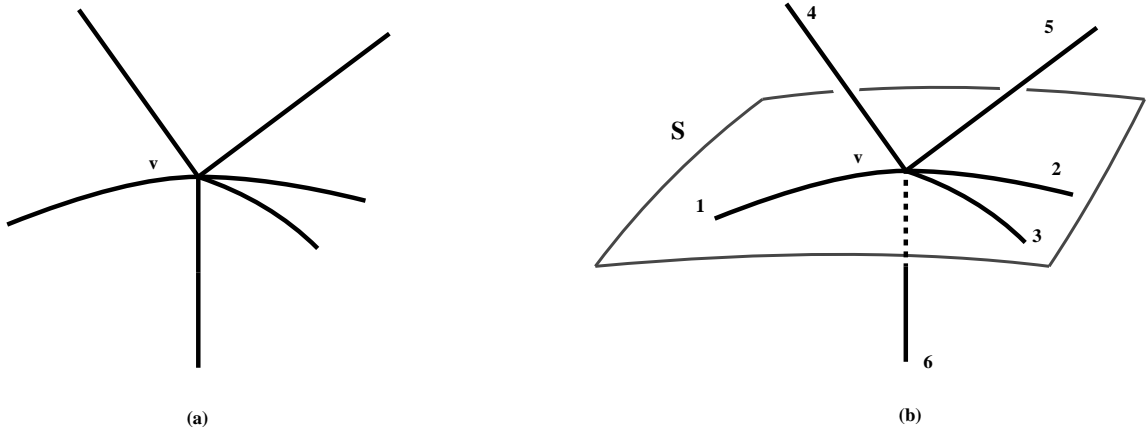}}}
\caption{Vertex of a 3-d spin network (a) and its intersection with the surface S (b). Edges $1,2,3$ are  tangential ones.}
\label{fig:1}
\end{figure}

\begin{figure}
\centerline{\hbox{\psfig{figure=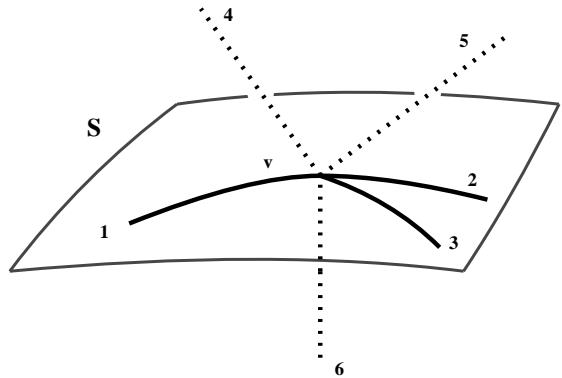}}}
\caption{Intertwining operator assigned to the vertex $v$ of a generalized surface spin network `remembers' what were the spins $j_4,j_5,j_6$ of the edges $4,5,6$ incident at the surface.}
\label{fig:2}
\end{figure}

Let us now specify the space of quantum surface states. Given a 2-d (not necessarily closed) surface $S$ embedded into the spatial manifold 
$\Sigma$ and a 3-d spin network $\Gamma$ 
one can consider the intersection 
of this spin network with $S$. Let us call the intersection 
points vertices. Generally, there can be vertices of any 
valence not less then two\footnote{%
In the case of theory without fermionic degrees of freedom, which we 
consider here, the valence of vertices of a spin network state 
should be not less than two in 
order to have a gauge invariant state. When fermionic degrees of 
freedom are present in the the theory valence of vertices can be 
equal to one. In this case open ends of a spin network describe fermionic degrees of freedom.} 
(valence of a vertex is the number of edges of a spin network state 
which meet in this vertex). Also, there can be
edges lying entirely on the surface $S$ among the edges of $\Gamma$ ; we shall call such edges 
tangential (see Fig. \ref{fig:1}). 
The intersection of $\Gamma$ with the surface $S$ defines 
what we shall call a surface spin network on $S$. A surface spin network is a graph lying entirely on the surface $S$, with spins assigned to each edge, and intertwining operators assigned to each vertex. The intertwiners assigned to each vertex are just those of the corresponding 3-d spin network. This means that vertices of a 2-d surface spin network `remember' what were the spins of the edges incident at the surface (see Fig. \ref{fig:2}). Note that our definition of a surface spin network is not the canonical one. A `canonical' surface spin network is defined as a graph lying on the surface, with spins assigned to each edge, and intertwiners assigned to each vertex. We use the term `surface spin network' simply to denote the intersection of a 3-d spin network with the surface $S$, or, in other words, to denote the `surface' part of information carried by a 3-d spin network. To avoid confusion, we shall also use the term `generalized surface spin network state'. 

\begin{figure}
\centerline{\hbox{\psfig{figure=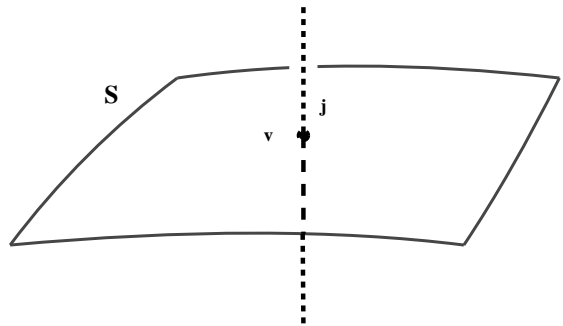}}}
\caption{The valence of the simplest vertex is two.}
\label{fig:3}
\end{figure}

The simples non-trivial example of a surface spin network is that coming from a single edge intersecting the surface $S$ (see Fig. \ref{fig:3}).
Such a spin network is simply the point $v$ on $S$, with the intertwining operator being the map from the one copy of the representation space $\rho^{(j)}$ ($j$ here is the spin labeling the irreducible representation of $SU(2)$) to the other copy of $\rho^{(j)}$. The intertwiner in this case is specified (up to an overall constant) simply by the spin $j$ attached to the vertex $v$.

We define the space $\cal H$ of 
surface quantum states as the space spanned by all (generalized) surface 
spin network states. This definition  means that the basis in $\cal H$ is formed by surface spin network states, what gives us all that we need for our counting purposes. 

We can now recall that there exists a set of well defined operators 
${\hat A}_R$, which `measure' quantum geometry of $S$. These operators correspond to areas of various regions $R$ of the surface 
$S$ (see \cite{QGeometry}). It turns out that 3-d spin 
network states are eigenstates of operators ${\hat A}_R$. The 
corresponding eigenvalues are given by
\begin{equation}
A_{s} = \sum_{v\in R} {1\over2} \sqrt{2j_{(v)}^{d}(j_{(v)}^d+1)+
2j_{(v)}^{u}(j_{(v)}^u+1)-
j_{(v)}^{u+d}(j_{(v)}^{u+d}+1)}.
\label{qarea}
\end{equation}
Note that we measure areas in the units of $16\pi l_p^2$, which is convenient in the loop quantum gravity.
Here the sum is taken over all vertices $v$ lying in the region $R$ 
of the surface $S$;
$j_{(v)}^u,j_{(v)}^d$ and $j_{(v)}^t$ are the total spins of edges 
lying up, down the surface and tangential to the surface (see 
\cite{QGeometry} for details). Although these operators are defined on 3-d quantum states, the eigenvalues (\ref{qarea}) depend only on the `surface' part of the information carried by a 3-d spin network state. Therefore, we can think of $\hat{A}_R$ as operators defined on the (generalized) surface spin network states, with eigenvalues given by (\ref{qarea}).

As we have said in the introduction, in this paper we are going to consider a geometrical entropy that corresponds to a macro-state specified simply by a total surface area $A$. For this simple case, the approximation criterion between macro- and micro-descriptions is straightforward. Namely, we can say that a quantum state (micro-state) approximates a given macro-state if the mean value of the operator ${\hat A}_S$ in this quantum state is approximately equal to the fixed value $A$. It is straightforward to see that quantum states approximating total surface area form a linear subspace in the space of all surface states. Its dimension is equal to the number of different surface spin network states approximating $A$.

Let us see now how many different quantum states 
approximate a given total area. It is easy to see that this number is infinite.
Indeed, loops on  the surface, which are the simplest possible parts of a surface spin network state, do not give any
contribution to the areas. Therefore, one has an infinite number of 
spin networks which approximate one and the same total area of $\cal S$
being different only in configurations of loops on the surface. 

One can argue, however, that this happens because, using terminology from statistical mechanics, a macro-state of our system is not completely specified when we fix only a total area of the surface. Indeed, areas of regions on $S$ carry information only about degrees of freedom described by $e^{AB}$ field. But there are also degrees of freedom described by the pull-back of the connection field on the surface which one should take care of when specifying a macro-state. Our guess is that different configurations of loops on the surface from the above example correspond to different classical configurations of the connection field on $S$. Indeed, as we know, in classical theory loop quantities are constructed as traced holonomies of the connection, and, therefore, are just those objects which carry information about the connection field. Therefore, since we want to forget about degrees of freedom described by the field $a^{AB}$ when we specify a macro-state, to be consistent we have also to forget about those quantum states which, as we believe, contribute to $a^{AB}$ and do not contribute to the area of $S$.

Let us, therefore, consider only quantum states which contribute to the areas of regions on $S$, but do not `contribute' to the connection field on the surface. These, according to our guess, are the states which contain no loops on the surface.
We shall call the corresponding spin networks {\it open} spin networks. A (generalized) surface spin network is called open if the (surface) graph that labels it contains no closed paths (or no loops). The simplest open spin network state is the spin network containing a single vertex (see Fig. \ref{fig:3}).

Now, to find the entropy which corresponds to a macro-state of a fixed total surface area $A$ we should calculate the number of quantum states which approximate $A$, taking into account only open spin networks. However, let us first analyze the problem taking into account only some particularly simple spin network states, and then try to generalize the result obtained.

\section{Geometrical entropy: sets of punctures.}
\label{sec:3}

\begin{figure}
\centerline{\hbox{\psfig{figure=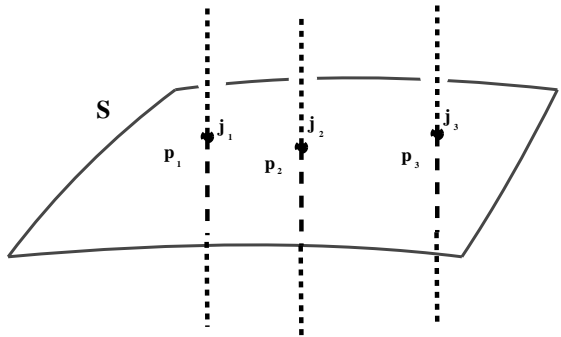}}}
\caption{The simplest surface spin network states are specified by a set of punctures on the boundary.}
\label{fig:4}
\end{figure}

Let us consider spin networks whose vertices are bivilent, i.e., those specified simply by sets of points (vertices) on the surface with spins assigned to these  points (see Fig. \ref{fig:4}).
Points on the surface with spins assigned are sometimes called punctures (see, for example, \cite{Linking}), and we shall use this name as well. So quantum states which we consider now are specified by sets of punctures on the surface.

For these simple quantum states there exists a simplified version of the formula (\ref{qarea}). Namely, a set $\{p,j_{p}\}$
of punctures gives the area of a region $R$ on $S$
\begin{equation}
A(\{p,j_{p}\}) = \sum_{p\in R} \sqrt{j_{p}(j_{p}+1)}.
\label{areasimple}
\end{equation}
Here the sum is taken over all punctures which lie in the region $R$ and $j_p$ are the corresponding spins. To get the total area of the surface we just have to sum over all punctures $p\in S$.

With these simple states the problem of calculating the entropy becomes almost straightforward.  

Let us use the standard trick. Instead of counting states which 
correspond to the same area we shall take a sum over {\it all} states
but take them with different statistical weight. Namely, let us consider
the sum
\begin{equation}
Q(\alpha) = \sum_{\Gamma}\exp{\left ( -\,\alpha\,A(\Gamma) \right )}
\label{StatSum}
\end{equation}
over all different states $\Gamma = \{p,j_{p}\}$, where $\alpha > 0$ is a
parameter. Considering $p_{\Gamma} = {1\over Q(\alpha)} \exp{\left ( -\,\alpha\,A(\Gamma) \right )}$ as a {\it probability}
of our system to be found in a state $\Gamma$, it is easy to see that the
{\it mean} value of the area in such statistical state is
\begin{equation}
A(\alpha) = - {\partial \ln Q\over\partial\alpha}.
\label{a1}
\end{equation}
The entropy of the system in this macro-state is given by the standard formula
$S = -\,\sum_{\Gamma} p_{\Gamma}\ln p_{\Gamma}$ or, as it is easy to check, by
\begin{equation}
S(\alpha) = \alpha\,A(\alpha) + \ln Q(\alpha).
\label{SA}
\end{equation}

We see that the mean value of the surface area depends on $\alpha$. If statistical `ensemble' of states is chosen properly, then one can
adjust the value of $\alpha$ in such a way that $A(\alpha)$ acquires any 
prescribed value. There is some particular  value of $S$ which corresponds to 
the chosen value of $A$. Excluding $\alpha$ in such a way we obtain the 
entropy as a function of the area $S = S(A)$. Statistical mechanics tells 
us that when the density $\eta(A)$ of states of our system grows sufficiently fast with $A$ it is of no difference which way of calculating $S(A)$ to choose; one can
count the logarithm of the number of different states which give one and the same area or calculate the function $S(A)$ as described - the results will not
differ. Let us note that $S(A)$ calculated through (\ref{SA}) will be meaningful only for
large $A$ (as compared with unity, i.e. with the Planckian area). This is
really what we need because only in this case the notion of approximation 
and, therefore, the notion of entropy acquires sense.

Let us now discuss whether all the sets of $\{p,j_{p}\}$ 
should be taken into account in (\ref{StatSum}).
First of all, let us recall that we have to count not the surface spin networks themselves but  the diffeomorphism equivalence classes of spin networks. This means that two surface spin networks which can be transformed on into another by a diffeomorphism on the surface should be considered as a single state in (\ref{StatSum}). Thus, the continuous set of data $\{ p, j_p \}$ ($p$ runs all over the surface $S$) reduces simply to a set of spins $\{ j_p \}$ when one identifies sets of punctures which can be mapped one into another by a diffeomorphism on the surface (note, however, the discussion following the next paragraph).  

Next, we note that, if the surface $S$ is a closed one, not every set $\{ p, j_p \}$ can be obtained as a result of intersection with $S$ of some 3-d spin network in $\Sigma$.  Namely, the sum $\sum_{p}j_{p}$,
which is the total spin which `enter' the surface must be an integer for gauge invariant states.
This corresponds to the fact that not all eigenvalues given by the formula (\ref{qarea}) are eigenvalues of the area of a closed surface, as it was first pointed out in \cite{QGeometry}. Let us consider in this section only a simpler case of an open surface $S$; we return to the case of a closed surface in sec. \ref{sec:4}.

\begin{figure}
\centerline{\hbox{\psfig{figure=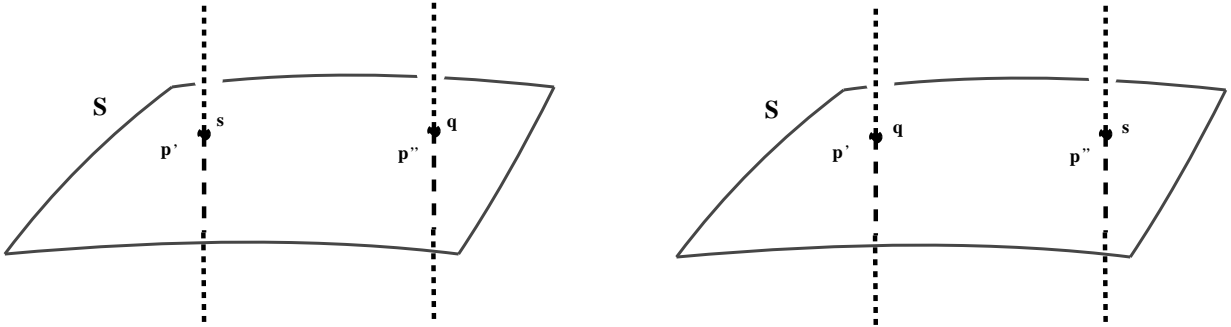}}}
\caption{Two sets of punctures which are considered as specifying different quantum states.}
\label{fig:5}
\end{figure}

And, finally, let us consider the states specified by the following two set of punctures (see Fig. \ref{fig:5})
\begin{equation}
\{\ldots,j(p')=s,\ldots,j(p'')=q,\ldots\},\qquad
\{\ldots,j(p')=q,\ldots,j(p'')=s,\ldots\}.
\label{states}
\end{equation}
These two states differ one from another only at two points $p',p''$; they give the same total area of the boundary. Should we distinguish these 
two states or consider them as a single physical state? 
It turns out that this is the key question which determines the form of the dependence of the entropy $S(A)$ on the area $A$ of the surface. Let us now consider these states as {\it different} and postpone the discussion of such a choice to the last section.
So let us denote by $N(A)$ the number of states which correspond to one and the same area $A$ in the case when all the sets (\ref{states}) are regarded as 
specifying different states. It is easy to see that $N(A)$ is the number of {\it ordered} sets of punctures which approximate the total area $A$. It is straightforward to compute $N(A)$ using our method with
the statistical sum.\footnote{%
One can also do this calculation explicitly, using combinatorial methods. See
\cite{Rov}.}
One can easily check that the fact that we regard the sets (\ref{states}) as specifying different states (or, equivalently, the fact that we count ordered sets of punctures) means that we can sum over the spin of each puncture {\it independently} 
\begin{equation}
Q = 1 + \sum_{n=1}^{\infty}\sum_{j_{p_{1}}=1/2}^{\infty}\cdots
\sum_{j_{p_{n}}=1/2}^{\infty}\exp{-\,\alpha\,
\sum_{p}\sqrt{j_{p}(j_{p}+1)}}.
\label{1}
\end{equation}
The first sum here denotes the sum over the number of possible punctures on $S$ and the subsequent ones denote the summation over the possible spins. It is easy to see that 
\begin{equation}
Q = {1\over 1 - z(\alpha)},
\label{2}
\end{equation}
where $z(\alpha)$ is given by
\begin{equation}
z(\alpha) = \sum_{j=1/2}^{\infty}\exp{-\,\alpha\,\sqrt{j(j+1)}}.
\label{smallz}
\end{equation}
Note that the sum here runs over all positive integers and half-integers.
One can expect that $Q(\alpha)$ will increase as $\alpha$ gets smaller because
in any case (\ref{1}) diverges when $\alpha$ goes to zero. However, we see
that (\ref{2}) diverges even for some finite value of $\alpha'$ such that 
$z(\alpha') = 1$ (we shall see in a minute to which value of $\alpha$ this 
corresponds). When $\alpha$ gets smaller and approaches $\alpha'$, $Q(\alpha)$
increases and diverges at the point $\alpha'$. It can easily be checked that
$A(\alpha)$  and $S(\alpha)$ also diverge when $\alpha \to \alpha'$. This
means that changing $\alpha$ slightly we will obtain substantial 
differences
in values of $A$ and $S$. What we are interested in is the dependence $S(A)$
for large values of $A$. But we see that all large values of $A$ can be
obtained by small changes in $\alpha$ when $\alpha \to \alpha'$. Thus, from
(\ref{SA}) we conclude that for $A >> 1$ 
\begin{equation}
S \approx \alpha'\,A.
\label{entropy}
\end{equation}
Here we neglected the term $\ln Q$ which is small comparatively with the main
term (\ref{entropy}).

This result tells us that entropy grows precisely as the first power of area
of the boundary. Now let us see what $\alpha'$ amounts to. One can do this
numerically but it is also straightforward to find an approximate value. Note
that $z(\alpha)$ can be rewritten as the sum over integers
\begin{equation}
z(\alpha) = \sum_{l=1}^{\infty} \exp{-\,{\alpha\over 2}\,\sqrt{l^2+2l}}.
\end{equation}
The term under the square root in the exponential can be given the
form $(l+1)^{2}-1$. Because the sum starts from $l=1$ we can neglect the unity
comparatively to a larger term. Then $z(\alpha)$ can be easily computed 
\begin{equation}
z(\alpha) = {\exp{(-\alpha)}\over 1\,-\,\exp{(-\alpha/2)}}.
\label{z}
\end{equation}
This gives for the $\alpha': z(\alpha')=1$ the value 
$\alpha'=2\,\log{{2\over\sqrt{5}-1}} \approx 0.96$. An explicit 
numerical investigation of the equation $z(\alpha') = 1$ gives a close value $\alpha'\approx 1.01$.

So we have shown that for large values of $A$ the entropy depends on the 
area as
\begin{eqnarray}
S(A) = \alpha' A; \nonumber \\
\alpha' = 1.01.
\label{formula}
\end{eqnarray}

\section{Geometrical entropy: the case of closed surfaces.}
\label{sec:4}

As we have mentioned before, in the case of closed surfaces ${\cal S}$ we have to take into account the fact that not all eigenvalues given by the formula (\ref{qarea}) are eigenvalues of the operator $\hat{A}_S$, which measures the total area of the surface. Namely, in the case of a closed surface ${\cal S}$ gauge invariant quantum states are those which satisfy the condition that the sums $\sum_{v\in S} j_{(v)}^u$ and $\sum_{v\in S} j_{(v)}^d$ over all vertices lying on the surface are integers \cite{QGeometry} (spins $j_{(v)}^u,j_{(v)}^d$ are those defined after the formula (\ref{qarea}). This condition means that for the case of a closed surface some spin network states should be excluded from the sum (\ref{StatSum}). Recall that our result essentially follows from the fact that the statistical sum $Q(\alpha)$ diverges when $\alpha$ approaches some {\it finite} value $\alpha'$. Because we have to drop some positive terms from the statistical sum corresponding to a closed ${\cal S}$, $Q(\alpha)$ can prove to be 
convergent for all $\alpha > 0$ and as a result we would obtain some other 
(non-linear) dependence $S(A)$\footnote{%
In fact, in this case we would obtain that S(A) grows {\it slower} than $A$.}.

The aim of this section is to show that geometrical entropy $S(A)$ of (ordered) sets of punctures, considered in the previous section, for a closed surface $S$ still depends linearly on the total surface area. So, again, our states are specified by ordered sets $\{p,j_p\}$ of punctures on the surface. However, in the case of a closed surface, not all sets of punctures correspond to gauge invariant physical states. Namely, in our case gauge invariant states are those for which $\sum_{p\in S} j_p$ is an integer. To find the number of different states which approximate one and the same total area we can again apply our trick with the statistical sum. The statistical sum $Q(\alpha)$ will be given by the expression (\ref{1}), where we have to take into account only the states satisfying the above condition.

It is not hard to calculate the statistical sum $Q(\alpha)$ for our case of a closed surface $S$. Let us divide the function $z(\alpha)$ given by (\ref{z}) into two parts $z(\alpha)=\tilde{z}(\alpha)+\tilde{\tilde{z}}(\alpha)$. Function $\tilde{z}(\alpha)$ is the sum over all integer values $l$ of spin $j$
\begin{equation}
\tilde{z}=\sum_{l=1}^{\infty} \exp{-\alpha\sqrt{l(l+1)}}.
\end{equation}
Function $\tilde{\tilde{z}}(\alpha)$ is the sum over all half-integers $j=l-1/2$
\begin{equation}
\tilde{\tilde{z}}(\alpha) = \sum_{l=1}^{\infty} \exp{-\alpha\sqrt{(l-1/2)(l+1/2)}}.
\end{equation}
Let us rewrite the statistical sum $Q(\alpha)$ over all sets of punctures in terms of the functions $\tilde{z},\tilde{\tilde{z}}$ introduced
\begin{equation}
Q(\alpha) = {1\over 1-(\tilde{z}+\tilde{\tilde{z}})} = 
{1\over 1-\tilde{z}} \sum_{n=0}^{\infty} \left({\tilde{\tilde{z}}\over 1-\tilde{z}} \right)^n.
\label{q1}
\end{equation}
A moment of reflection shows that in order to get the statistical sum corresponding to the case of a closed surface, we have to drop from (\ref{q1}) all terms which contain odd powers of $\tilde{\tilde{z}}$. Thus, we get
\begin{equation}
Q(\alpha)_{closed} = {1\over 1-\tilde{z}} \sum_{n=0}^{\infty} \left({\tilde{\tilde{z}}\over 1-\tilde{z}} \right)^{2n} = {1\over 1-\tilde{z}} \;
{1\over 1-\left({\tilde{\tilde{z}}\over 1-\tilde{z}}\right)^2}.
\end{equation}
Here $\tilde{z},\tilde{\tilde{z}}$ are functions of $\alpha$. Recall now that our statistical mechanical system has a regime in which the entropy depends linearly on the area in the case when the statistical sum diverges for some finite value of $\alpha$. Let us, therefore, investigate the behavior of $Q(\alpha)_{closed}$ when $\alpha$ goes to zero. First, let us note that
\begin{equation}
Q(\alpha)_{closed} = {1-\tilde{z}\over (1-\tilde{z}+\tilde{\tilde{z}})(1-\tilde{z}-\tilde{\tilde{z}})}.
\end{equation}
Also, we note that $\tilde{\tilde{z}} > \tilde{z}$, because the sum in $\tilde{\tilde{z}}(\alpha)$ starts from $j=1/2$, whereas the sum in $\tilde{z}$ is taken over all integer values of spin and starts from $j=1$. We see, therefore, that the statistical sum $Q(\alpha)_{closed}$ diverges for $\alpha': \tilde{z}(\alpha')+\tilde{\tilde{z}}(\alpha') = 1$. But $\tilde{z}+\tilde{\tilde{z}}=z$, so the value of $\alpha'$ here is the solution of equation $z(\alpha') = 1$ obtained in the previous section.

Thus, we get that
\begin{equation}
S_{closed}(A) = \alpha' \,A,
\end{equation}
where $\alpha'$ is the same as in (\ref{formula}). Thus, we have proved that, although the statistical sum $Q_{closed}(\alpha)$ for the case of a closed surface is different from $Q(\alpha)$ corresponding to the case of an open $S$, it implies the linear dependence $S(A) = \alpha' A$ of the entropy on the surface area, with the proportionality coefficient $\alpha'$ being the same as in the case of an open surface. One can say that, although we excluded some states from the statistical sum, there are still `enough' states to give the same linear dependence of the entropy on $A$ as in the case of an open surface.

\section{Discussion.}

The result which we have obtained considering some particularly simple surface quantum states is that the entropy corresponding to a macro-state which is specified by a total area $A$ of the surface is proportional precisely to the area $A$. It is important that we have obtained precisely the same dependence $S(A)$ both for the case of open and closed surfaces $S$. The entropy was defined as the logarithm of the number of different quantum states that approximate one and the same area of the surface. The states which we considered were specified simply by the sets of punctures on the surface. It is crucial that the states which are different only up to a permutation of spins (see Fig.\ref{fig:5}) were considered as different quantum states.

One of the reasons why we postponed the discussion of the key point of distinguishing the states (\ref{states}) was to emphasize the importance of this choice. This is this choice which implies the linear dependence of entropy $S(A)$ on the surface area. As we show in the appendix, in the case when states (\ref{states}) are considered as indistinguishable, the dependence of the entropy on the areas is different form the linear one (in fact, the entropy turns out to be proportional to the square root of the area).

Before discussing this key point, let us note that our aim in this paper is not to give a physical motivation for some particular choice of the ensemble of quantum states. Rather we wanted to show how the entropy arises naturally when one considers correspondence  between classical and quantum pictures of geometry. We also wanted to illustrate this idea following some simple choice of the ensemble of quantum states, and to present a technique which turns out to be useful.  Having this in mind, let us discuss our choice of  considering the states (\ref{states}) as distinguishable ones. 

We were counting the number of different diffeomorphism equivalence classes of (simple) surface spin networks approximating one and the same total surface area. This means, that two surface spin networks which can be transformed one into another by a diffeomorphism on the surface should have been considered  as specifying the same micro-state. Let us now consider the two states (\ref{states}). It is easy to see, that there exists a diffeomorphism which maps one state into the other; this diffeomorphism simply rearranges two points $p',p''$ on the surface. Therefore, on the first sight, the states (\ref{states}) should have been considered as a single state.

Does this mean that the result obtained is physically meaningless and is simply an exercise in statistical mechanics?. There are some reasons to believe that the result obtained is more than that.  So let us give some possible motivations for our strange choice of considering states (\ref{states}) as different quantum states. 
 
One possibility, which is argued by Carlo Rovelli \cite{Rov}, is that points on the surface are physically distinguishable, and so are the states (\ref{states}). This, in fact, happens in the case of some systems for which the surface $S$ play the role of the boundary. It might be the case that {\it boundary conditions} partly (or even completely) brake diffeomorphism invariance on the boundary. This would mean that some spin networks which are usually considered as specifying one and the same quantum state, should, in fact, be considered as different quantum states. This case of systems with boundaries is subtle and deserves a special attention. Let us only note the possible connection of our result, viewed from this point, with the results of Steven Carlip \cite{Carlip}.

The other possibility, which has also been argued in \cite{SM}, is that some other (in fact, loop) states on the surface make some of the states (\ref{states}) belong to different diffeomorphism equivalence classes of spin networks. Indeed, the surface loop states which we considered as giving no contribution to the areas, and, thus, forgot about, may affect largely the number of different diffeomorphism equivalence classes of states which approximate one and the same area of $S$. To see this, let us introduce a loop configuration on the surface. This loop configuration divides the surface into regions. It is clear that some of the states (\ref{states}) will belong to different equivalence classes, for there will no longer be a diffeomorphism `connecting' different regions. Thus, the number of different diffeomorphism equivalence classes which approximate one and the same total area $A$ in this case is larger that in the case when there are no loops on the surface. So loops on the surface may allow one to distinguish states of the form (\ref{states}) (for more details see \cite{SM}).

Of course, neither of these motivations gives a final physical justification of the result obtained. But let us repeat that this is not what we aimed at in this paper. We hope, however, that  the above discussion shows at least that the issue deserves a further investigation. 

\begin{figure}
\centerline{\hbox{\psfig{figure=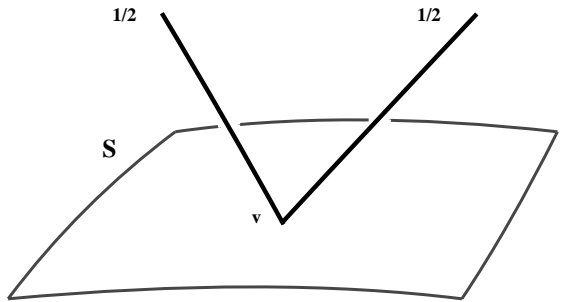}}}
\caption{Surface state which does not contribute to the surface area, thus, producing  the degeneracy.}
\label{fig:7}
\end{figure}

Finally, let us discuss the possibility to generalize the result obtained considering arbitrary open surface spin network states. First, let us take into account surface spin networks which have no tangential edges, allowing, however, vertices of arbitrary valence. In this case we have to use a general formula (\ref{qarea}) for eigenstates of area operators (with all $j_{(v)}^{(u+d)}$ being equal to zero because of the fact that we consider spin networks with no tangential edges). We would like to generalize our result counting all (open) surface spin networks which approximate one and the same total surface area. However, we face the problem trying to consider all states. Namely, eigenvalues given by the formula (\ref{qarea}) are degenerate and we have to take this degeneracy into account when calculating the entropy. Let us consider, for example, a simple state, which contains one vertex of valence two (see Fig. \ref{fig:7}). 
For this state we get $j^u = j^d = 0$, and, therefore, $A_S = 0$. A moment of consideration shows that there are, in fact, an infinite number of similar surface states that do not give any contribution to the area of $S$. Thus, we find that eigenvalue $A_S=0$ is infinitely degenerate. Similarly, we find that {\it all} eigenvalues given by the formula (\ref{qarea}) are infinitely degenerate. Therefore, if we would like to take into account all different surface states we would get an infinite value for our geometrical entropy. 
   
Let us note, however, that the states which we have just considered are rather pathological. Namely, we observe that small deformations of the surface $S$ (see Fig. \ref{fig:7}) (with the spin network state being not deformed) cause a change in the `quantum area' of $S$. Let us consider a one parameter family of surfaces $S_\epsilon, \epsilon\in [0,1]$ such that $S_\epsilon \to S$ when $\epsilon\to 0$. For our example (see Fig. \ref{fig:7}), if surfaces $S_\epsilon$ approach the surface $S$ from below we have $\lim_{\epsilon\to 0} A_{S_\epsilon} = A_S = 0$ (here by $A_S$ we denote an eigenvalue of the operator $\hat{A}_S$ on the quantum state we consider). However, if we choose the family of surfaces approaching $S$ from above we have $\lim_{\epsilon\to 0} A_{S_\epsilon} = \sqrt{3} \not= A_S = 0$ (we measure areas in the units $16\pi l_p^2$). Thus, we see that states which cause the degeneracy are `pathological' when considered as surface states, for `quantum area' of $S$ in this states behaves non-continually under small deformations of $S$. This observation, which the author learned from A.Ashtekar, suggests that we have to exclude these states when we consider an ensemble of surface states. A natural way to do it would be to change the approximation criterion between macro- and micro-descriptions. Namely, let us strengthen our criterion in the following way. We fix a value $A$ of the total area of $S$, which defines our macro-state. We choose one parameter family $S_\epsilon$ of two-surfaces such that $S_\epsilon\to S, \epsilon\to 0$. Let us now say that a spin network state $\Gamma$ approximates our macro-state if $\lim_{\epsilon\to 0} A_{S_\epsilon}(\Gamma) = A$, where $A_{S_\epsilon}(\Gamma)$ is the eigenvalue of operator $\hat{A}_{S_\epsilon}$ corresponding to the eigenstate $\Gamma$. 

The new approximation criterion states that we have to consider only `good' surface quantum states. By definition, `quantum area' of $S$ in `good' surface quantum states does not change under small deformations of the surface $S$. It is easy to see, that `good' quantum states are those which have only bivilent vertices, i.e. precisely those states which we considered in this paper. Thus, we conclude that the result obtained above gives the geometrical entropy $S(A)$ of a macro-state of a fixed total area $A$, the quantum states that account for this entropy being all states which approximate $A$ in the strong sense, i.e., those for which `quantum area' does not change under small deformations of the surface $S$. Furthermore, the entropy $S(A)$ is the same both for open and for closed surfaces. Thus, the result obtained is general in the sense that we consider all quantum surface states which approximate $A$ in the strong sense. 

Let us conclude saying that the notion of geometrical entropy is, presumably,
valid not only in the form explored here (when we fixed only one macroscopic
parameter -- the surface area), but also in a more general context. For
example, it is of interest to calculate the entropy $S(g)$ which corresponds
to a given 2-metric on the surface, which is a genuine geometrical entropy.

\section{Acknowledgments.}

I am grateful to Yuri Shtanov for the discussions in which the idea of 
geometrical entropy developed and for important comments on the first 
versions of the manuscript. 
I would like to thank A.Ashtekar, A.Coricci, R.Borisov, S.Major, C.Rovelli, L.Smolin and J.Zapata for discussions,
comments and criticism. I am grateful to A.Ashtekar from whom I learned the idea of sequences of surfaces, which is used here to formulate the `strong' approximation criterion.
I  would also like to thank the Banach center of
Polish Academy of Sciences for the hospitality during the period when this
paper was started. This work was supported, in part by the International
Soros Science Education Program (ISSEP) through grant No. PSU062052.

\appendix
\section{Counting unordered sets of punctures.}

In this appendix we would like to show that the entropy $S(A)$ in the case when one considers sets (\ref{states}) as undistinguishable is proportional to the square root of the area. 

Since we do not distinguish between
sets (\ref{states}), all states $\Gamma$ which enter (\ref{StatSum})
are just unordered sets $\{j_1,j_2,\ldots\}$ of spins. One can describe 
each such state by a set $\{n_{j_1},n_{j_2},\ldots\}$ of numbers of spins 
$j_1,j_2,\ldots$ in this state. It is convenient to imagine each state 
defined by a set $\{j_1,j_2,\ldots\}$ as a collection of different
particles. Particles of a particular sort $j$ correspond to punctures
carrying the spin $j$. There can be many particles of the same sort in a
state and we denote the corresponding number of such particles by $n_j$. 
Having this analogy in
mind it is easy to see that our case the statistical sum $Q(\alpha)$ is given by 
\begin{equation}
Q(\alpha) = \sum_{\{n_j\}} \exp{\left ( -\alpha\sum_{j=1/2}^{\infty} n_j A(j)\right )}.
\label{s2}
\end{equation}
Here by $A(j)$ we denoted the contribution to the area from a 
single `particle' of the sort $j$
\begin{equation}
A(j) = \sqrt{j(j+1)}.
\label{area3}
\end{equation}
Note that the sum over $j$ starts from $j=1/2$ and runs over all positive integers and half-integers.
It is straightforward to take the sum over numbers $n_j$ 
of different `particles' in (\ref{s2}); this gives
\begin{equation}
Q(\alpha) = \prod_j {1\over 1-\exp{(-\,\alpha A(j)})}.
\end{equation}
The area as a function of parameter $\alpha$ is given by the formula
(\ref{a1}). In our case it gives
\begin{equation}
A(\alpha) = \sum_j {A(j)\,\exp{(-\,\alpha A(j))}\over 1-\exp{(-\,\alpha A(j))}}.
\label{3}
\end{equation}
To see what dependence of $A$ on $\alpha$ this implies we replace the sum over spins $j$ by the sum over integers (colors) $l=2j$. Then $A(l) = {1\over 2} \sqrt{l^2+2l}$. For values of $l>1$ we have the following approximation 
\begin{equation}
A(l) \approx (l+1)/2.
\label{approx}
\end{equation}
Let us now approximate the sum over $j$ in (\ref{3}) by the following integral over $l$
\begin{equation}
A(\alpha) \approx \int_{l=1}^{\infty} dl 
{{(l+1)\over2}\,\exp{-\,\alpha {(l+1)\over2}}
\over 1-\exp{-\,\alpha {(l+1)\over2}}}.
\end{equation}
It is easy to see that this integral is equal to
\begin{equation}
A(\alpha) = {2\over\alpha^2} \int_{\alpha}^{\infty} dx 
{x\,\exp{-x}\over 1-\exp{-x}}.
\label{4}
\end{equation}
Recall now that we are interested in the dependence $S(A)$ for large values
of $A$ (as compared with unity). We see from (\ref{4}) that large values
of $A$ correspond to small $\alpha$. In the limit $\alpha\to0$
the integral in (\ref{4}) can easily be calculated
\begin{equation}
\int_0^\infty dx {x\,\exp{-x}\over 1-\exp{-x}} = {\pi^2\over6}.
\end{equation}
So, finally, for large values of $A$ (or, equivalently, for small $\alpha$),
we have the following dependence 
\begin{equation}
A(\alpha) = {\pi^2\over3\alpha^2}.
\end{equation}
This gives the dependence $S(A)$ of the entropy on the area. Namely, in the limit of large $A$ we can neglect the logarithmic term in (\ref{SA}) as compared with the larger first term. Thus, we get
\begin{equation}
S = {\pi\over\sqrt{3}}\,\sqrt{A}.
\label{entropy1}
\end{equation}
So, in the case when one considers unordered sets of punctures, the geometrical entropy $S(A)$ is proportional to the
square root of the area.


\begin{thebibliography}{99}
\bibitem{QGeometry} A.Ashtekar, J.Lewandowski, Quantum theory of 
geometry I: Area operators, preprint gr-qc/9602046.

\bibitem{GeomEig} R.De Pietri, C.Rovelli, 
Geometry Eigenvalues and Scalar Product from Recoupling Theory in
Loop Quantum Gravity, preprint gr-qc/9602023.

\bibitem{SpinNet} C.Rovelli, L.Smolin, Phys.Rev.D {\bf 53}, 5743 
(1995).

\bibitem{Weave} A.Ashtekar, C.Rovelli, L.Smolin, Phys.Rev.Lett., {\bf 69},
237 (1992).

\bibitem{Asht1} A.Ashtekar, J.Lewandowski, D.Marolf, J.Mourao and T.Thiemann,
J.Math.Phys. {\bf 36}, 6456 (1995).

\bibitem{Linking} L.Smolin, Linking non-perturbative quantum gravity and topological field theory, preprint gr-qc/9505028.

\bibitem{Rov} C.Rovelli, Black Hole Entropy from Loop Quantum Gravity,
preprint gr-qc/9603063

\bibitem{Carlip} S.Carlip, Statistical mechanics and black hole entropy, preprint gr-qc/9509024.

\bibitem{SM} K.Krasnov, On statistical mechanics of gravitational systems,
preprint gr-qc/9605047.

\end{thebibliography}
\end{document}